# 2001 QR322: a dynamically unstable Neptune Trojan?

J. Horner[1*] and P. S. Lykawka[2#]
[1] Department of Physics and Astronomy, The Open University, Walton Hall, Milton Keynes, MK7 6AA, UK
[2] International Center for Human Sciences (Planetary Sciences), Kinki University, 3-4-1 Kowakae, Higashiosaka, Osaka, 577-8502, Japan

---

[*] E-mail address: j.a.horner@open.ac.uk; Current address: Department of Physics, Science Laboratories, Durham University, South Road, Durham, DH1 3LE, United Kingdom
[#] Previous address: Department of Earth and Planetary Sciences, Kobe University, 1-1 rokkodai-cho, nada-ku, Kobe 657-8501, Japan.

**ABSTRACT**

Since early work on the stability of the first Neptunian Trojan, 2001 QR322, suggested that it was a dynamically stable, primordial body, it has been assumed this applies to both that object, and its more recently discovered brethren. However, it seems that things are no longer so clear cut. In this work, we present the results of detailed dynamical simulations of the orbital behaviour of 2001 QR322. Using an ephemeris for the object that has significantly improved since earlier works, we follow the evolution of 19683 test particles, placed on orbits within the observational error ellipse of 2001 QR322's orbit, for a period of 1 Gyr. We find that majority of these "clones" of 2001 QR322 are dynamically unstable, exhibiting a near-exponential decay from both the Neptunian Trojan cloud (decay halflife ~550 Myr) and the Solar system (decay halflife ~590 Myr). The stability of the object within Neptune's Trojan cloud is found to be strongly dependent on the initial semi-major axis used, with those objects located at $a \geq 30.30$ AU being significantly less stable than those interior to this value, as a result of their having initial libration amplitudes very close to a critical threshold dividing regular and irregular motion, located at ~70-75º (full extent of angular motion). This result suggests that, if 2001 QR322 *is* a primordial Neptunian Trojan, it must be a representative of a population that was once significantly larger than that we see today, and adds weight to the idea that the Neptune Trojans may represent a significant source of objects moving on unstable orbits between the giant planets (the Centaurs).

# 1 INTRODUCTION

The Jupiter Trojans, such as asteroid 588 Achilles, have been known and well studied for over a hundred years. It is currently thought that they may well outnumber the objects in the asteroid belt (Yoshida & Nakamura 2005), having moved on their stable 1:1 mean-motion resonant orbits since the birth of the Solar system. Less than a decade ago, the first Neptunian Trojan was discovered. This object, 2001 QR322 (hereafter, 'QR322', for brevity), and the five other Neptune Trojans discovered since, are thought to be the first members of a population believed to be even larger than that of the Jupiter Trojans – icy relics held in deep freeze since our Solar system was formed (Chiang et al. 2003; Sheppard & Trujillo 2006).

In the first few years following its discovery, some study was made of the dynamical behaviour of QR322. Chiang et al. (2003) integrated a small number of clones of QR322, on orbits derived from the object's observational range of orbital elements, for 1 Gyr. They found that the test particles displayed stable behaviour over that period. Marzari, Tricarico & Scholl (2003) used dynamical simulations to conclude that there is a good chance the object is primordial in nature, with just 7 out of 70 clones followed for 4.5 Gyr escaping from the Trojan cloud by the end of their simulations. Also in 2004, based on the simulation of 100 clones of QR322, Brasser et al. found that, although the dynamical behaviour of the object within the Trojan cloud was complex, due to the effects of the $\nu_{18}$ nodal secular resonance (characterised by the libration of $\Omega - \Omega_N$ with time, where $\Omega$ is the longitude of the ascending node and the subscript $_N$ refers to Neptune), most of its clones remained stable in integrations spanning the last 5 Gyr. Care should be taken when using the results of these early works to predict the behaviour of 2001 QR322, since they were based on derived orbital elements for the object based on only a short period of observation. In the time since those projects were carried out, the observation arc available to calculate the current orbit of the object has grown from just a couple of years to almost a decade. As a result, the nominal orbit for the object has changed somewhat. In addition, little information is given on how the clones of QR322 were created, and insufficient detail is provided regarding the settings used for the integrations of Marzari et al. (2003). Also, the orbital uncertainties of the orbit used for QR322 are not given in either Marzari et al. (2003) or Brasser et al. (2004). It is therefore unclear unclear how much store should be set in their results.

In recent years, the object has been somewhat ignored, its stability assumed on the basis of these past works, as an ever-growing number of objects in the outer Solar system have vied for attention. On the basis of some preliminary trials (detailed in Lykawka et al. 2009), we found that there is some evidence that QR322 is in fact somewhat dynamically unstable, capable of leaving the Neptunian Trojan cloud after as little as 19 Myr – far shorter than the Gyr lifetimes proposed by previous authors. These differences likely result mainly from the improvements made to the known orbit of QR322 – and as such, it seems that the time is right for a fresh look at this intriguing object.

Therefore, we revisit the dynamical stability of QR322 by following the orbits of almost 20,000 clones, distributed to cover the full 3σ orbital uncertainties in all six elements, for 1 Gyr. Given the short dynamical timescales described above, it is clear that simulations of this duration are ample to allow us to obtain a good statistical view of the behaviour of the object, and are short enough to enable us to improve our statistics through the simulation of as many test particles as possible. Our calculations incorporate significant improvements over previous works. First, we used orbital elements derived from longer-arc observations (taken in early 2009), which means that the elements will be more refined and more precise. In addition, the number of clones used in our simulations is approximately 200-300 times larger than in Marzari et al. (2003) and Brasser et al. (2004). This higher resolution allowed us to extend our study to examine the dependence of QR322's dynamical stability on its initial orbital inclination, and the three rotational orbital elements, which represents a significant improvement on these earlier works.

In Section 2 of this work, we describe the technique used to model the long-term dynamical behaviour of QR322 as a function of time, before presenting and discussing our results in Section 3, and drawing together our conclusions in Section 4.

**2 SIMULATING 2001 QR322**
Since previous work on the orbital behaviour of QR322 was carried out, the nominal orbit for the object has changed as a result of an increased number of observations being made. Though the inclination remains unchanged, the change in the other orbital elements means that these earlier works may now bear little or no relation to the behaviour of this particular object. The most recent orbit available for QR322 at the time our simulations began, obtained on 2009 January 26 from the AstDys website (http://hamilton.dm.unipi.it/astdys/), is given in Table 1.

| Element | Value | $1\sigma$ error |
|---|---|---|
| $a$ (au) | 30.3023 | 0.008813 |
| $e$ | 0.031121 | 0.0003059 |
| $i$ (°) | 1.323 | 0.0009417 |
| $\Omega$ (°) | 151.628 | 0.02328 |
| $\omega$ (°) | 160.73 | 0.8316 |
| $M$ (°) | 57.883 | 0.7818 |
| Epoch | 2454800.5 | |

**Table 1:** Orbital elements, together with the degree of uncertainty in those elements, of 2001 QR322 as listed on the AstDys website on 2009 January 26. These values are based on an observational arc of 1450 days, and will undoubtedly undergo small changes as further observations are made.

In order to carry out detailed dynamical simulations of the behaviour of QR322, a number of clones of the object were created in orbital element space, spreading out in all 6 elements to a distance of $3\sigma$ in each direction. Nine clones were made in each of $a$, $e$, and $i$, evenly spread across the $6\sigma$ range centred on the nominal value. The resulting initial distribution of the test particles in $a$-$e$-$i$ space is shown in Figure 1. In addition, for each $a$-$e$-$i$ value, 3 clones were made in each of the three rotational elements, each separated by $3\sigma$. Consequentially, each location within $a$-$e$-$i$ space (as shown in Fig. 1) was tested by a total of 27 different test particles ($3^3$ objects in $\Omega$, $\omega$ and $M$ per $a$-$e$-$i$ tested). In total, this cloning procedure therefore created 19683 clones ($9^3$ in $a$-$e$-$i$, times $3^3$ in $\Omega$-$\omega$-$M$) in a cloud centred upon the nominal orbit for the object, the central clone of which had the exact orbital parameters detailed in Table 1.

Using the *Hybrid* integrator contained within the MERCURY dynamics package (Chambers 1999), these clones were followed under the gravitational influence of the giant planets (Jupiter, Saturn, Uranus and Neptune) until they were either ejected from the Solar system, on reaching a distance of 1000 au from the Sun or collided with one of the massive bodies. The integrations followed the objects for a period of 1 Gyr, with orbital data output at 10,000 year intervals, and an integration time step of 1/3 of a year.

**3 RESULTS AND DISCUSSION**
Of the initial swarm of 19683 test particles, just 7220 survived for the full 1 Gyr of integration time. In other words, just over 63% of the particles were either ejected to a heliocentric distance of 1000 au, or collided with one of the massive bodies (the four giant planets and the Sun) over integrations covering less than one quarter of the accepted age of our Solar system. Of those objects which remained in the system at the end of the integrations, 6809 were still moving on orbits within the Neptunian 1:1 mean-motion resonance, with the remaining 411 objects having left that region onto more dynamically unstable orbits.

The left hand panels of Fig. 2 show the decay of the population of test particles over the duration of our integrations. The black line shows the number of particles remaining anywhere within the Solar system (i.e. all those which had not collided with a massive body or been ejected beyond 1000 au), while the red line shows those which remain on Trojan orbits. The upper plot shows the decay in these numbers as a function of time, while the lower contains the same data, plotted in log-log format. The right hand panels of Fig. 2 show a highly simplified illustration of the stability of the Trojan cloud, plotting the "Running Halflife" as a function of time. This value, which is intended to be a purely illustrative tool, is calculated using the equation

$$T_{1/2} = \frac{\ln 2 \, t}{\ln(N) - \ln(N_0)}, \qquad (1)$$

Where $T_{1/2}$ is the Running Halflife, $N_0$ is the size of the initial population (19683 objects), and $N$ is the number remaining after time $t$. In other words, therefore, it is the value of decay halflife calculated at the exact moment a given clone is removed from the system, assuming that the decay has been perfectly exponential to that point. From these plots, it is clear that, after an initial period of a few million years during which the clones of QR322 disperse throughout the Neptunian Trojan cloud, the number of objects surviving within the Trojan region (red line), and the system as a whole (black line), decays in a roughly exponential manner. As time goes on, the less stable clones are gradually removed, which leads to the "Running Halflife" gradually rising as only the most stable clones remain. However, the decay is close enough to being truly exponential that we can use that fact to calculate the dynamical halflife of QR322. Following Horner, Evans & Bailey (2004), we analysed the data with the help of least-square fitting routines from Press et al. (1992), which allowed us to calculate the dynamical halflife for the sample, both for removal from the Trojan cloud and removal from the system as a whole. For details of the maths behind our calculations we refer the interested reader to Horner et al. (2004).

| Criterion | Decay halflife (Myr) | Sigma on halflife (Myr) |
|---|---|---|
| Survival as a Neptune Trojan | 552.74 | 1.76 |
| Ejection from the Solar system | 593.01 | 1.43 |

**Table 2:** Dynamical halflife for the survival of 2001 QR322 as a Neptune Trojan (upper row) and as a Solar system object (lower row), following Horner et al. (2004).

Our results suggest that, although QR322 is a dynamically long-lived object (with a dynamical lifetime at least 1/8[th] the age of our Solar system), it can not be considered truly dynamically stable on Gyr timescales, as had previously been believed. Nevertheless, the length of the objects dynamical lifetime means that the possibility remains that the object could be a representative of a decaying population of primordial objects - particularly if such a population was captured during the migration of the giant planets, as discussed in Lykawka et al. (2009). It is, of course, true that the decay rate would likely decrease as time went on, as the least stable clones would be preferentially lost, and the most stable clones would be more likely to remain in the Trojan cloud. Nevertheless, even in the case where there is a significant spread in dynamical lifetimes across the cloud of clones, such a Trojan population would continue to decay through the entire lifetime of the Solar system, and that particular region of the Neptunian Trojan cloud would therefore be expected to contain only a small fraction of its initial population at the present epoch.

To illustrate the relative instability of the orbit of this object, and also highlight how the stability varies as a function of the orbital elements, we note that the mean lifetime of the 27 clones moving on orbits with the *a*, *e*, and *i* of the nominal orbit (a 3x3x3 cube in rotational elements) is approximately 300 Myr, significantly less than the overall mean dynamical halflife detailed above.

Figs 3, 4 and 5 show the relationship between the mean dynamical lifetime and the initial semi-major axis, eccentricity and inclination of the test particles. It is clear that there is a region of instability correlated with the initial semi-major axis of the test particles, which appears to be intriguingly relatively independent of the particles location in $e$-$i$-$\omega$-$\Omega$-$M$ space. In contrast, those particles lying toward the inner edge of the cluster, in semi-major axis space, showed far greater levels of dynamical stability.

The large area of instability stretching outward from 30.29 au is the primary reason for the dynamical instability of QR322, and deserves further discussion. In order to examine the source of this instability, we examined a number of representative clones for each of the different initial semi-major axis values used. Given that no dependence was found between the mean dynamical lifetime of the suite of test particles and their eccentricity or inclination (e.g. Figs 3-5), and no such dependence was apparent upon the three rotational orbital elements, $\omega$, $\Omega$ and $M$, we chose representative particles that were spread across a wide range of values in these elements for each value of semi-major axis examined. Firstly, we found that the libration amplitude of clones of QR322 tends to slightly increase toward larger values of initial semi-major axis. Clones that started within the stable/marginally stable regions with initial $a < 30.30$ au yielded libration amplitudes, $A$, of less than approximately 60-70º (the scale of the full angular motion of the object around the Lagrange point). In this region, clones which escaped after a period of several hundred Myr showed no secular trends or abrupt increases in eccentricity, inclination, or libration amplitude prior to leaving the Trojan cloud. Instead, the escapes were found to occur shortly after the objects libration amplitudes increased from roughly 65-70º to values in the range 70-75º. However, no apparent monotonic trend was observed in the dynamical evolution of these objects. Those clones which began life within the unstable region, with $a \geq 30.30$ au, displayed dynamical behaviour that was, superficially, remarkably similar to that described above for their more stable brethren. However, it was clear that the significantly shorter escape timescales observed for these objects was the direct result of their having typically slightly larger initial libration amplitudes, which allowed them to more rapidly evolve onto orbits approaching the critical threshold of $A = 70$-$75$º. In short, therefore, the large area of instability observed in Figs 3 and 4 is the direct result of the gradual increase of the initial libration amplitude of clones of QR322 as a function of semi-major axis. The closer the initial amplitude to a critical value of ~70-75º, the more rapidly the object can evolve into this unstable regime, and be ejected from the cloud.

So, once clones of QR322 evolve to orbits with $A \sim 70$-$75$º, their orbits become markedly less stable. What is the cause of this instability? Upon examination of the unstable clones, we found no evidence of any strong apsidal secular resonances between the objects and either Neptune or Uranus (the $\nu_8$ and $\nu_7$ resonances), aside, of course, from the $\nu_{18}$ resonance between all the representative clones and the planet Neptune. Similarly, no secular resonances were found between the objects and those two planets combined. We also checked for low-order combinations involving the longitudes of perihelion of the clones, Uranus, and the two gas giant planets, Jupiter and Saturn. Again, no libration or slow-circulation of such arguments was observed in the evolution of these clones. What, then, about the $\nu_{18}$ secular resonance? Despite the fact that it appears the only secular resonance which is capable of influencing the behaviour of QR322, we found that this resonance has a negligible effect on its dynamical evolution, a result supported by previous studies of the effect of the $\nu_{18}$ resonance on the evolution of low-$i$ Neptunian Trojans (e.g. Marzari et al. 2003; Zhou, Dvorak & Sun 2009). Indeed, the clones of QR322 studied in this work did not display any noticeable inclination excitation, in contrast to the findings of Brasser et al. (2004).

Taken together, then, all these results suggest that the apparent dynamical instability of QR322 is not linked to the effects of secular resonances on its orbit. It is clear, therefore, that some other mechanism must be acting to cause the instability. As discussed above, those clones of QR322 which have initial $A$ in the range 65-70º tend to become unstable well within our 1 Gyr integration timescale. It is interesting to note that such libration amplitudes overlap with the approximate boundary between chaotic and regular motion (where regular motion is likely to yield far more stable dynamic behaviour), according to dynamical diffusion maps for orbits around the Neptunian Lagrangian points (Nesvorny & Dones 2002; Marzari et al. 2003; Zhou et al. 2009). In particular, Zhou et al. (2009) built detailed dynamical maps in which Trojans experiencing such regular motion are found to confined to a region roughly delineated by a maximum $A$ of ~70º, for low-$i$-orbits ($i$ < 10º). This region is therefore postulated to be a region of long-term stability, a result which is supported by the results of direct Gyr-scale orbital integrations performed by these authors.

It therefore seems reasonable to conclude that the orbital solution for QR322 used in this work (Table 1) places the nominal orbit of the object, together with the clones used to examine its behaviour, right on the boundary between stable regular motion and unstable chaotic motion, albeit in a way that the majority of the clones manage to escape the Neptunian Trojan cloud within a period of 1 Gyr. The origin of the orbital instability seems to be related to a family of complex secondary resonances involving the frequencies of the Trojan librational motion, $f_{TR}$, the near 2:1 mean motion resonance between Uranus and Neptune, $f_{2N:1U}$, and the apsidal motion of Saturn, $g_6$ (Zhou et al. 2009). Indeed, close examination of the ejected clones revealed that a number showed near commensurabilities of the form $f_{2N:1U} - 2f_{TR} = g_6 - g$, where $g$ represents the frequency of the Trojan's apsidal motion.

This shows that QR322 is a fine example of the rich dynamics of Neptunian Trojans, and that further observations of this object are necessary in coming years in order to refine its orbit, before a definitive answer on its long-term stability can be reached. However, if QR322 is considered to be a typical Neptunian Trojan (or, at least, to be typical of the low inclination component of the Trojan family, with inclinations less than ~10°), then our results strongly suggest that, over the age of the Solar system, this population will have been continually supplying the outer Solar system with fresh dynamically unstable objects. Indeed, such an idea is well supported by our Fig. 2, in which the difference between the red and black lines reveals the ongoing existence of such a population through our integrations, amounting to between 5 and 10% of the population of QR322-like bodies at a given time, $t$. Although it is true that the slope of the decay rate may well flatten during the last 3 Gyr (as a result of the less-stable members of the population being preferentially lost), the escape of such objects may well still represent a significant contribution to the population of unstable bodies in the outer Solar system, particularly if the initial mass of the Neptunian Trojan cloud was significantly higher than the current value.

Analysis of the orbital distribution of the 412 survivors that did not lie on Trojan orbits at the end of the integrations reveals that these bodies had acquired orbits typical of both the Centaurs and the full gamut of trans-Neptunian objects (TNOs). Examples include particles moving on orbits similar to those of objects within the classical Edgeworth-Kuiper belt (dynamically cold orbits within ~50 au), resonant TNOs (trapped within the 3:2, 2:1 and more distance mean-motion resonances with Neptune), scattered TNOs (with a variety of eccentricities and inclinations, and perihelia within 40 au) and even detatched TNOs (objects with perihelion > 40 au) (see Lykawka & Mukai 2007a, 2007b and Gladman et al. 2008 for details about nomenclature). This result adds weight to the idea that the Neptunian Trojans may act as an important source of the Centaur population (and in turn, contribute a significant fraction of the objects within the Jupiter family of comets), and may also contribute to the captured Trojan populations of the other giant planets, as discussed in Horner & Lykawka (2010) and Lykawka & Horner (2010), respectively. In these papers, we detail further n-body integrations which reveal that significant fractions of both the clones of QR322 and theoretical Trojan objects (objects captured by, and transported with, Neptune, during its postulated migration outward during the final stages of planetary formation) that leave the Neptunian Trojan cloud evolve onto orbits that are indistinguishable from those of the known Centaurs, and that with only moderate assumptions, these escapees can contribute a significant fraction of the Centaur population (Horner & Lykawka 2010). We have also show than all four giant planets are able to both capture and retain a significant population of Trojan objects from a primordial trans-Neptunian disk during the process of planetary migration. Since it is well known that large populations of Saturnian and Uranian Trojans are unlikely to exist at the current day, and given that our results show that many of the captured Trojans of Jupiter and Neptune are trapped on orbits that are only moderately stable, the bulk of the objects captured by all four planets will have been released back to dynamically unstable orbits in the outer Solar system over the course of its evolution, providing an ongoing, and potentially large, contribution to the population of Centaurs (Lykawka & Horner 2010).

**4 CONCLUSIONS**
The results of highly detailed dynamical simulations of 2001 QR322 show that this object is potentially far less dynamically stable than had previously been thought. Indeed, the members of a swarm of 19683 test particles, spread within the 3σ observational error ellipse on the objects orbit decay from the Neptunian Trojan cloud in an approximately exponential manner, with a dynamical halflife of just 553 Myr. The survival of the clones within our Solar system as a whole also displays a similar decay, with a slightly longer dynamical halflife of 593 Myr. The difference between these two lifetimes, just 40 Myr, is a direct result of the fact that the post-Trojan evolution of an ejected particle is highly chaotic, and therefore typically significantly shorter than its lifetime as a Neptunian Trojan. Although such a decay rate is long enough that we cannot rule out the idea that QR322 is a primordial Neptunian Trojan, if we assume that QR322 is a fairly typical example of a modern Neptunian Trojan, it at least suggests that the initial Neptunian Trojan population was far higher than that observed at the current day. Such a dynamically unstable population is fully compatible with one formed as a by-product of giant planet migration towards the end of planetary formation (e.g. Lykawka et al. 2009).

Plots of the stability of QR322 as a function of initial semi-major axis show that the dynamical lifetime is strongly correlated with the initial orbit used. Test particles placed on orbits with semi-major axes greater than, or equal to, 30.30 au are significantly more unstable than those which begin inward of this value. This sharp change from stable to unstable orbits occurs as a result of the fact that the initial libration amplitudes of the clones of QR322 places them right on the boundary between stable and unstable regions of phase space. Any test particles which evolve to orbits with libration amplitudes of order 70º begin to be destabilised by complex secondary resonances between the Trojan motion and the dynamics of at least three planets (Saturn, Uranus and Neptune). Those objects which begin on orbits with greater initial semi-major axis also begin with slightly larger libration amplitudes, meaning that they have less distance to evolve in libration amplitude space before they can be affected by these complex perturbations, leading to a shorter typical destabilisation time. Further observations of QR322 over the coming years will certainly reduce the errors on the calculated orbit significantly, and help to resolve the question of whether it lies within the more stable region, or is truly a dynamically unstable object. In light of the uncertainties involved, it is clear that the QR322 is far from a typical "stable Neptunian Trojan", and merits a significant amount of further study from both observers and theorists.

**ACKNOWLEDGEMENTS**
PSL and JAH gratefully acknowledge financial support awarded by the Daiwa Anglo-Japanese Foundation and the Sasakawa Foundation, which proved vital in arranging an extended research visit by JAH to Kobe University. PSL appreciates the support of the COE program and the JSPS Fellowship, while JAH appreciates the ongoing support of STFC. We would also like to thank the referee of this paper, Ramon Brasser, who made a number of very helpful comments and suggestions.

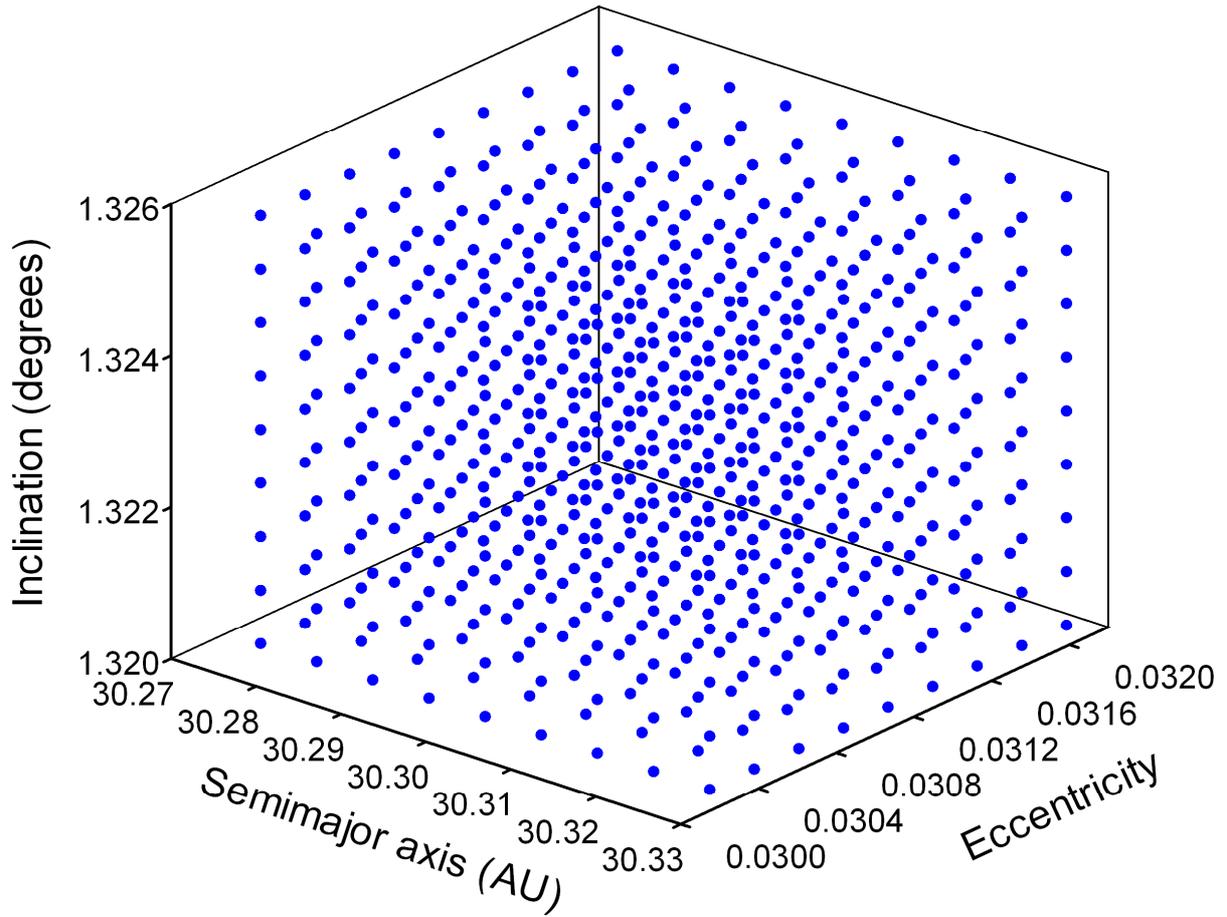

**Figure 1:** The initial distribution of test particles used to examine the dynamics of 2001 QR322, in *a-e-i* space. Detailed dynamical integrations were carried out using massless test particles on orbits spread across the region of *a-e-i* space shown, which is centred on the nominal orbit of QR322 as listed on the AstDys website on January 26[th], 2009. The test particles were spread out in each orbital element so that the outliers lay at a distance of 3σ from the nominal orbit. The cube of test particles therefore measures 6σ on a side, and has the nominal orbit of the object located at its exact centre. For each value of *a-e-i* space examined, a spread of 27 clones was tested, with a 3x3x3 cube distributed in the same manner in *Ω-ω-M* (so that, for example, test particles were placed at *ω - 3σ*, *ω*, and *ω + 3σ*). In this manner, a total population of 19683 clones were created, spanning $9^3$ locations in *a-e-i* space and $3^3$ locations in *Ω-ω-M* space.

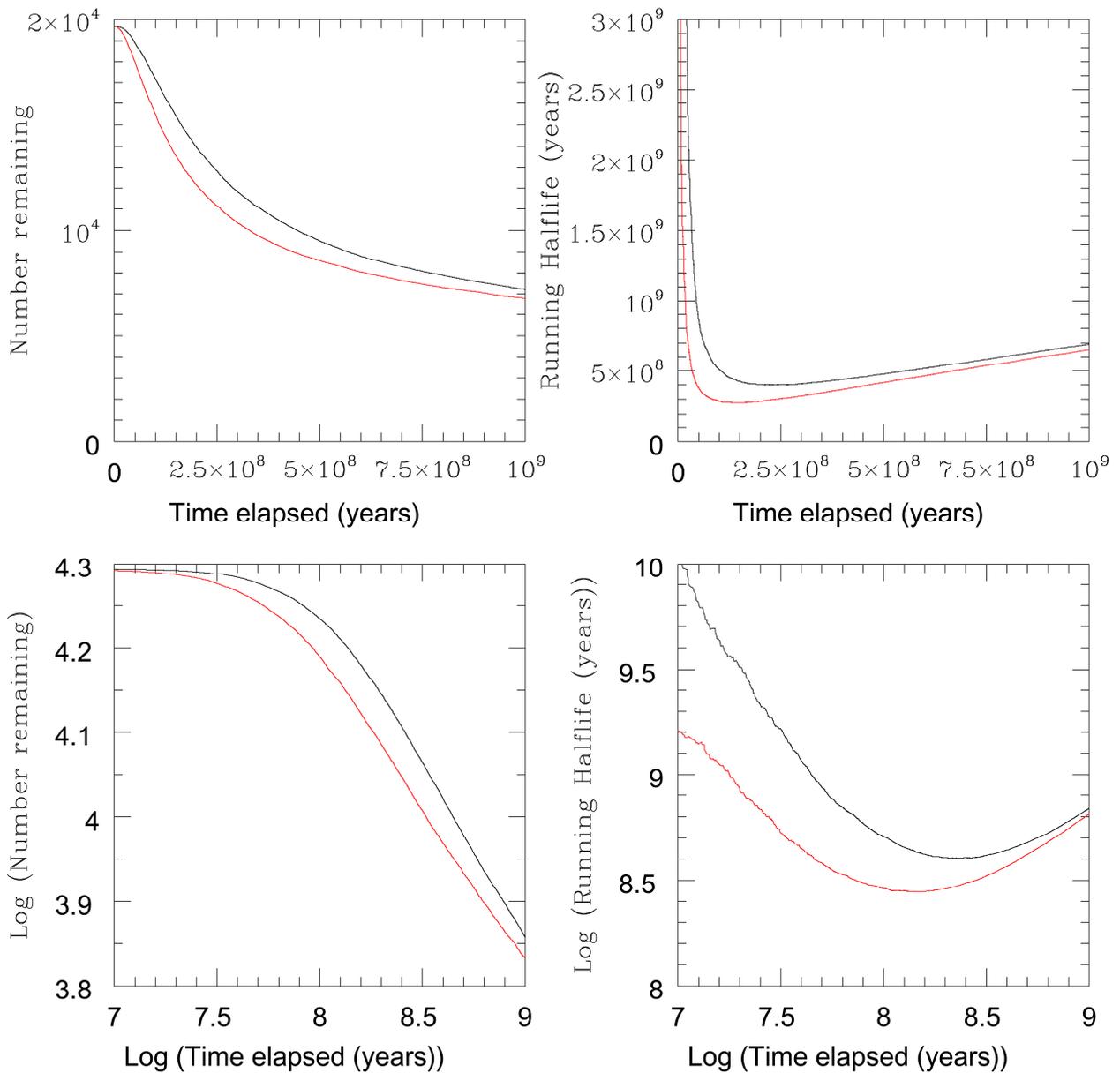

**Figure 2:** The left hand panels show the evolution of the population of clones of 2001 QR322 as a function of time, with the red line showing the number of objects remaining in the Neptunian Trojan cloud, and the black line the number remaining within the Solar system. The uppermost panel shows the decay as a function of time, while the lower panel shows the log of the number of surviving objects against that of the elapsed time, in years. It is clear that the decay is, to a first approximation, exponential in nature, albeit with a delayed start as the clones take time to diffuse from their initial locations. To illustrate this, the right hand plots show the "Running Halflife" (as detailed in the main text) for the population as a function of time. The initial slow dispersal of the clones (in the first few million years) can be clearly seen by the initially large values of this parameter, while the difference in decay rates from the Trojan region (red line) and Solar system (black line) can be clearly seen.

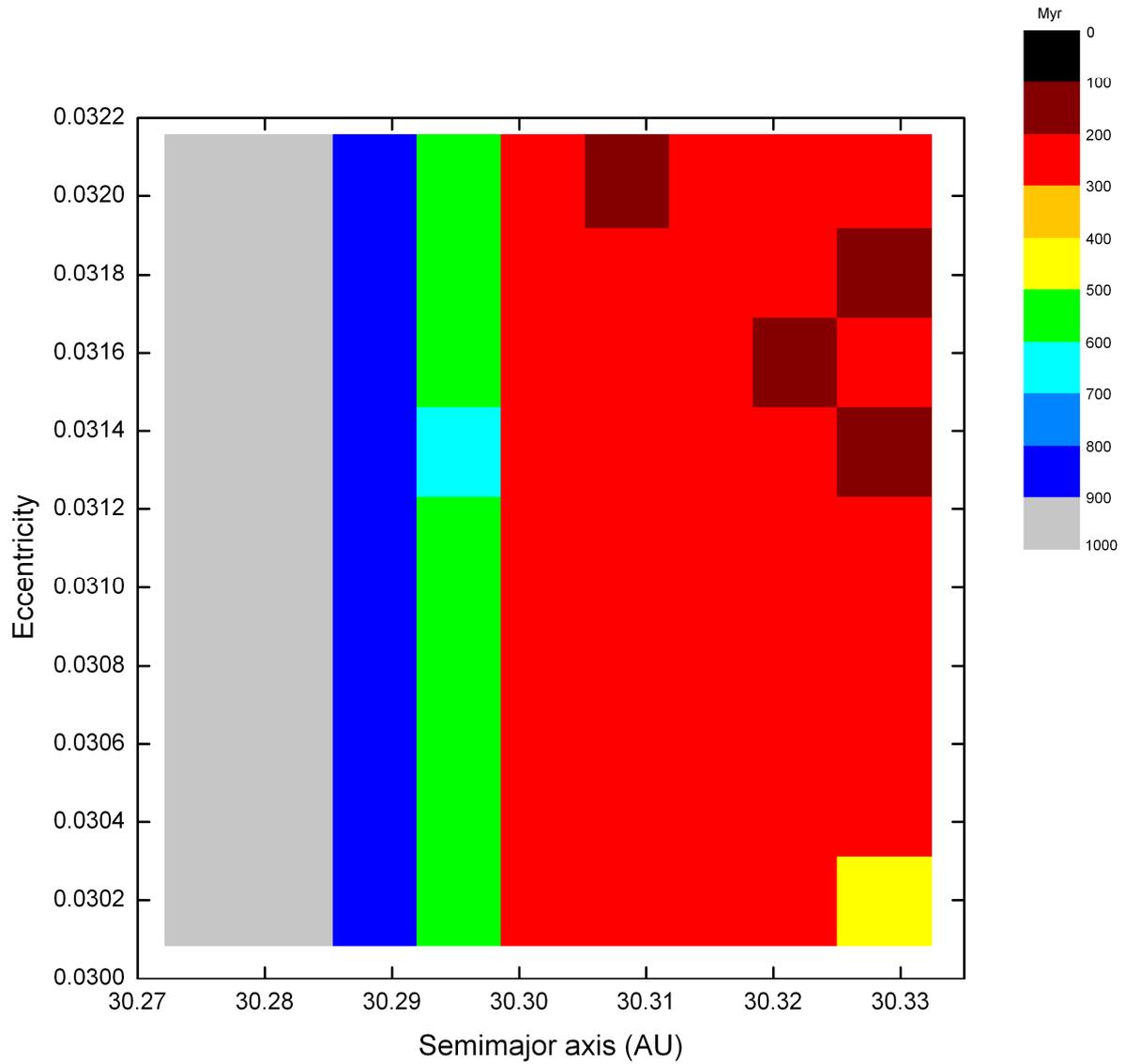

**Figure 3:** The mean lifetime of the clones of 2001 QR322 as a function of their initial semi-major axis and eccentricity. Note that the displayed orbital elements refer to a ±3σ distribution of particles, centred on the nominal orbit of the object at epoch JD 2454800.5 (as detailed in Table 1). Each square within the plot contains the combined results from 243 individual clones, spread over a variety of inclinations and rotation angles (9x3x3x3 – equivalent to a "block" of nine points in Fig. 1). The colour shows the mean dynamical lifetime of each bin, in 100 Myr increments. Note the strong dependence of dynamical stability on the initial semi-major axis of the particles.

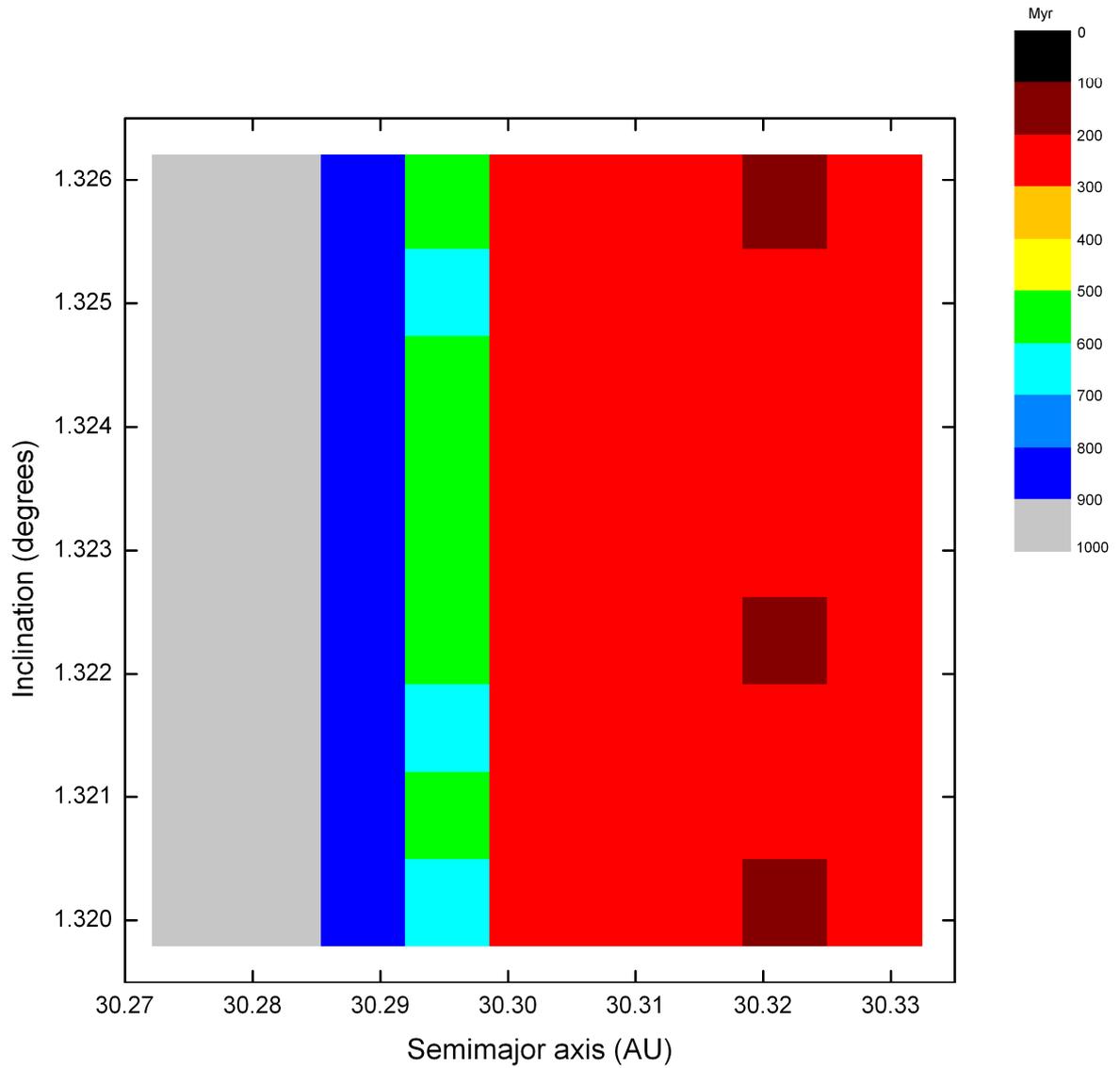

**Figure 4:** The mean lifetime of the clones of 2001 QR322 as a function of their initial semi-major axis and inclination. Note that the displayed orbital elements refer to a ±3σ distribution of particles, centred on the nominal orbit of the object at epoch JD 2454800.5 (as detailed in Table 1). Each square within the plot contains the combined results from 243 individual clones, spread over a variety of eccentricities and rotation angles (9x3x3x3 – equivalent to a "block" of nine points in Fig. 1). The colour shows the mean dynamical lifetime of each bin, in 100 Myr increments. Again, the strong dependence of dynamical stability on the initial semi-major axis of the particles can be clearly seen.

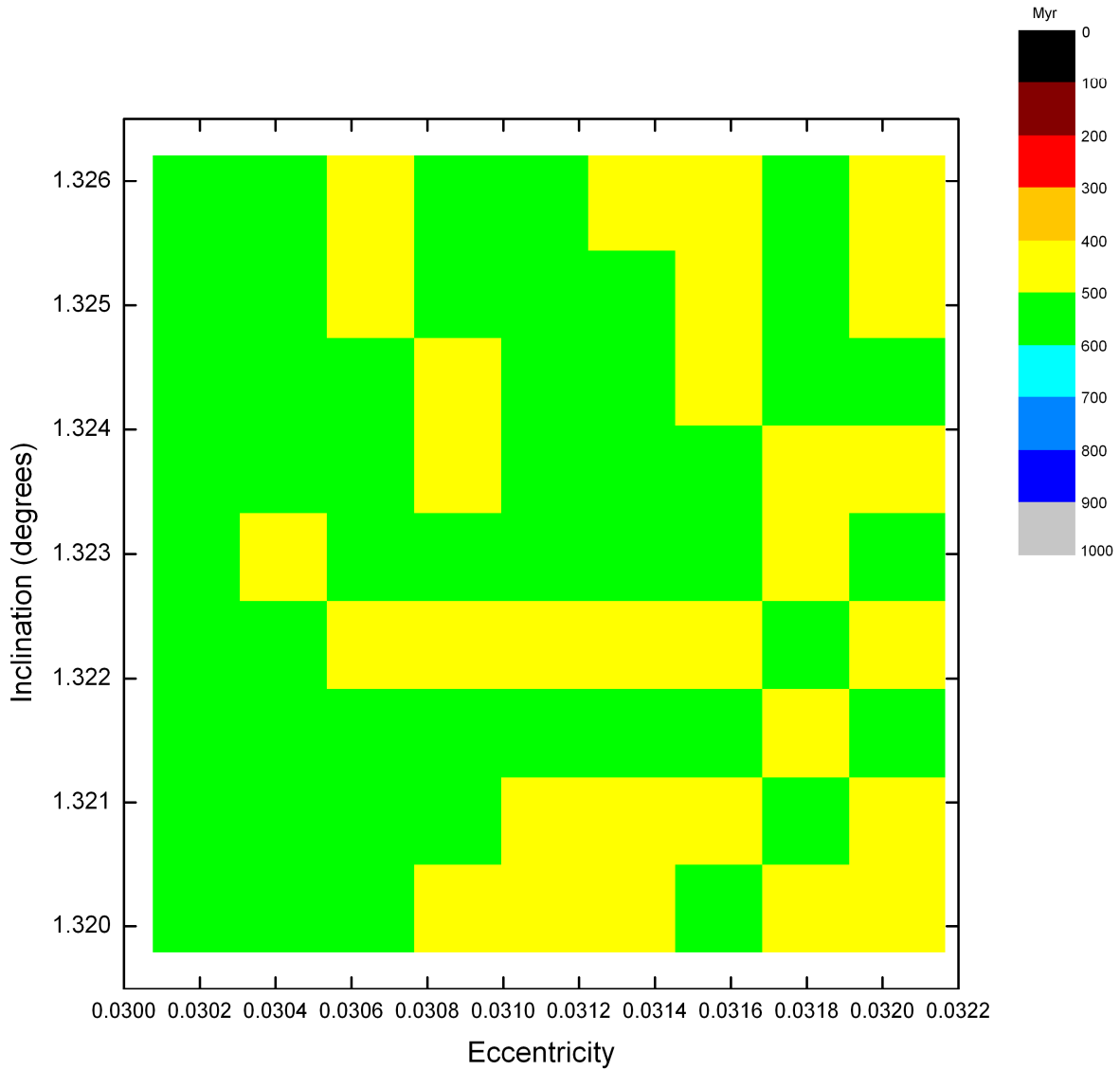

**Figure 5:** The mean lifetime of the clones of 2001 QR322 as a function of their initial eccentricity and inclination. Note that the displayed orbital elements refer to a ±3σ distribution of particles, centred on the nominal orbit of the object at epoch JD 2454800.5 (as detailed in Table 1). Each square within the plot contains the combined results from 243 individual clones, spread over a variety of semi-major axes and rotation angles (9x3x3x3 – equivalent to a "block" of nine points in Fig. 1). The colour shows the mean dynamical lifetime of each bin, in 100 Myr increments. No strong correlation between lifetime and inclination or eccentricity can be seen, although there are hints that objects on more eccentric orbits are the least stable, particularly when on orbits of lower inclination.